\documentstyle{article}

\newcommand{\pp}[1]{\phantom{#1}}
\newcommand{\be}{\begin{eqnarray}}
\newcommand{\ee}{\end{eqnarray}}
\newcommand{\ve}{\varepsilon}
\newcommand{\vs}{\varsigma}

\title{
Darboux-integrable equations with non-Abelian nonlinearities
}
\author{Nikolai~V.~Ustinov$^{1,2}$ and Marek Czachor$^{2,3}$\\
$^1$ Theoretical Physics Department, Kaliningrad State University, \\
Al.\,Nevsky street 14, 236041, Kaliningrad, Russia\\
$^2$ Katedra Fizyki Teoretycznej i Metod Matematycznych\\
Politechnika Gda\'{n}ska,
ul. Narutowicza 11/12, 80-952 Gda\'{n}sk, Poland\\
$^3$ Department of Physics\\
Technische Universit\"at Clausthal, 38678 Clausthal-Zellerfeld, Germany}

\begin{document}

\maketitle

\begin{abstract}
We introduce a new class of nonlinear equations
admitting a representation in terms of Darboux-covariant
compatibility conditions. Their special cases are, in particular,
(i) the ``general" von Neumann equation $i \dot \rho=[H,f(\rho)]$, with
$[f(\rho),\rho]=0$, (ii) its generalization involving certain functions
$f(\rho)$ which are non-Abelian in the sense that
$[f(\rho),\rho]\neq0$, and (iii) the Nahm equations.
\end{abstract}

\section{Introduction}

An investigation of collective phenomena in quantum mechanics leads
to various nonlinear evolution equations.
Nonlinear equation of a Schr\"odinger type was derived as a phenomenological
equation for the order parameter in superfluid He$_4$ \cite{Gross,Pitaevskii}.
Recent experiments on Bose-Einstein condensation \cite{exp}
significantly raise an interest in nonlinear generalizations of the
Schr\"odinger equation (for a review see \cite{DGPS}).
Another kind of nonlinear Schr\"odinger equations, a by-product of
work on classification of groups of diffeomorphisms \cite{DG}, was recently
related to certain aspects of D-brane dynamics \cite{Mavromatos}.

In more realistic situations, where entanglement between interacting
particles
is properly taken into account, one does not arrive at nonlinear
Schr\"odinger
wave equations but rather at their density matrix (von Neumann-type)
nonlinear
versions \cite{nuclear}
\be\label{Heis}
-i\dot X=[X,h(X)].
\ee
In this case the Hamiltonian $h(X)$ is considered as a ``non-Abelian function"
of the density operator $X$.
As is well known, equations analogous to (\ref{Heis}) are often
encountered in
quantum optics and field theory if one deals with the Heisenberg-picture
evolution of observables.

Still another class of nonlinear von Neumann type equations may be derived in
dissipative contexts \cite{Messer,Korsh} or on the basis of various
entropic variational principles \cite{Beretta,G-S}.

Of some interest is the fact that for a special class of $h(X)$
Eq.~(\ref{Heis}) can be rewritten
as
\be\label{vNeu}
i\dot X=[H,f(X)].
\ee
Nonlinear equations of this type appeared in the frameworks of nonlinear
Nambu-type theories \cite{MCpla97} and nonextensive statistics \cite{CN}.
It should be stressed that $f(X)$ does not always take the usual
form known from spectral theory \cite{Thirring}.
We will refer to the equations of the general form
(\ref{Heis}), (\ref{vNeu}) as nonlinear equations of the von~Neumann type.
Eq.~(\ref{vNeu}) acquires an additional fundamental flavor if one
recalls that
for $X=|\psi\rangle\langle\psi|$ and for {\it all\/} functions
constructed via
the spectral theorem, which satisfy $f(0)=0$ and $f(1)=1$, one
finds $f(X)=X$ and therefore the dynamics of $X$ is equivalent to the {\it
linear\/} Schr\"odinger equation.

Similar nonlinear equations can be found
also in classical theories.
The best known physical example is the Euler equation for
a freely rotating rigid body
\be\label{EA}
\dot X=[H,X^2].
\ee
The more abstract versions are related to the Euler-Arnold
equation for an ``$N$-dimensional rigid body" \cite{A}, the
Lie-Poisson equations
occurring in fluid dynamics \cite{A,MR}, and the $N$-wave equations
for electromagnetic waves in nonlinear media \cite{ZM}.

Particularly interesting and in recent years very intensively
investigated class of nonlinear equations are
the Nahm equations \cite{Hitchin}.  Their solutions are used as an
intermediate step in construction of non-Abelian monopoles.
One family of
solutions is in a one-to-one relationship to the Euler rigid-body
equations. In this sense the Nahm equations may be regarded as a kind
of generalized von Neumann equations.

Finally, quite recently nonlinear equations on free associative
algebras, including
the ones of the form (\ref{Heis}), (\ref{vNeu}) with $h(X)$ and $f(X)$ being
(noncommutative) polynomials, were considered in the framework of the
symmetry approach to classification of integrable ordinary
differential equations \cite{MS}.
It was found in particular that Eq.~(\ref{vNeu}), where $f(X)=iX^3$, is
symmetry for Eq.~(\ref{EA}).
A class of equations discussed in \cite{MS} was termed
``non-$C$-integrable". Below we show that some of them belong to our
class of integrable equations with Darboux covariant Lax representation.
It is worth mentioning that we do not assume a polynomial structure
of the RHS of Eqs.~(\ref{Heis}), (\ref{vNeu}).

As we can see, the reasons for generalizations of the linear von~Neumann
equation may be different, but they all finally lead to the same
fundamental difficulty:
The resulting equations involve a large (often infinite) number of degrees of
freedom and effective integration procedures are difficult to find.
The situation is additionally complicated by constraints typical of density
matrices or Hamiltonians.

It was only recently that soliton methods were applied to the density matrix
version of (\ref{vNeu}) \cite{SLMC,MKMCSL,ULCK}.
The progress was made possible by the observation that there exist
Darboux-covariant Lax representations of certain von Neumann-type equations.
The technique used in \cite{SLMC,MKMCSL,ULCK} is an appropriate modification
of the dressing method \cite{NZ,Levi,Mikh,Neu,Cieslinski} or rather
of its analogue
constructed via a binary Darboux transformation \cite{LU,U,L}.
The technique is called the Darboux transformation since
the construction of generalized gauge transformation is performed
with the help of additional solutions of the Lax pairs.

The Darboux-type method of integrating the density-matrix analogue of
Eq.~(\ref{EA}), introduced in \cite{SLMC} and further generalized in
\cite{ULCK}, led to discovery of the so-called self-scattering solutions.
The process of self-scattering continuously interpolates between two
asymptotically linear evolutions.
This very characteristic property was found in all the nontrivial solutions of
nonlinear von Neumann equation obtained by the above technique.

The paper presents further development of our results.
All the equations we derive can be regarded as
compatibility conditions for Darboux-covariant Lax pairs.
Previously discussed nonlinear von
Neumann equations as well as the Nahm equations are particular
examples of the class under consideration, but they form just a tip
of an iceberg.

Darboux covariance of Lax pairs is proved in detail along the lines of
\cite{nlin}. An alternative proof, involving a more general class of
Darboux transformations, is given in this volume in
\cite{Cieslinski'}. The two constructions are based on different
mathematical techniques, show different aspects of
the same problem, and it is not completely clear whether they are
entirely equivalent.

The layout of the paper is as follows.
The compatibility-condition representation of a family of nonlinear equations
of the von~Neumann type is given in Sec.~II.
The compatibility conditions are brought into a closed form by a
special choice of operator coefficients of the Lax pair.
The coefficients are defined in terms of additional functions
satisfying restrictions following from the
compatibility conditions.
A wide class of such functions is proposed.
Examples of nonlinear equations of the von~Neumann type that are generated
by some of these functions are given in Sec.~III.
Darboux covariance of the Lax pairs with the operator-valued
coefficients is proved in Sec.~IV.
In the next section we show that the restrictions carrying the compatibility
condition into integrable nonlinear von Neumann equations
are Darboux-covariant if the
functions introduced in Sec.~II are used.

\section{Darboux-integrable equations}

We begin with the overdetermined system of linear equations (the Lax pair)
\be\label{Lax}
\left\{
\begin{array}{rcl}
-i\dot\psi&=&\psi A(\lambda)\\
z_{\lambda}\psi&=&\psi H(\lambda)
\end{array}
\right. ,
\ee
where $\lambda$ and $z_{\lambda}$ are complex numbers, $\psi$ is an
element of a
linear space ${\bf L}$, $A(\lambda)$ and $H(\lambda)$ are linear operators
${\bf L}\mapsto{\bf L}$ belonging to an associative ring, the dot denotes a
derivative (i.e. an operator satisfying the Leibnitz rule).
The compatibility condition for the Lax pair is
\be
i\dot H(\lambda)=[A(\lambda),H(\lambda)]. \label{comp-cond}
\ee
Assume the operators entering the Lax pair are rational functions of
$\lambda$ with operator coefficients
\be\label{ABC}
A(\lambda)
&=&\sum_{k=0}^L\lambda^kB_k+\sum_{k=1}^M\frac{1}{\lambda^k}C_k,
\\
\label{H}
H(\lambda)&=&\sum_{k=0}^N\lambda^kH_k.
\ee
The compatibility condition implies two sets of relations between
operators $B_k$, $C_k$ and $H_k$
\be\label{setB}
\sum_{k=\max\{0,m-L\}}^{N}[H_k,B_{m-k}]&=&0\qquad(N<m\le L+N),
\\
\label{setC}
\sum_{k=0}^{\min\{N,m+M\}}[H_k,C_{k-m}]&=&0\qquad(-M\le m<0)
\ee
and the system of differential equations
\be\label{ode}
-i\dot H_m=\sum_{k=\max\{0,m-L\}}^{m}[H_k,B_{m-k}]+
\sum_{k=m+1}^{\min\{N,m+M\}}[H_k,C_{k-m}]
\ee
for $(0\le m\le N)$. In order to reduce Eqs.~(\ref{ode}) to equations
of the von~Neumann
type one needs to write them in a closed form.
In general, Eqs.~(\ref{setB}), (\ref{setC}) are inconvenient for defining
operators $B_k$ and $C_k$ in terms of $H_k$.
Nevertheless, one can express $B_k$ and $C_k$ explicitly through operator
$H_k$ by imposing on them some additional relations which obey
Eqs.~(\ref{setB}), (\ref{setC}).
It is clear that not all such additional relations have to
be consistent with the requirement of Darboux-covariance of the Lax
pair.

Consider
\be\label{B}
B_k
&=&
\frac{1}{(L-k)!}\left.\left(\frac{d^{L-k}}{d\vs^{L-k}}
f(\vs^NH(\vs^{-1}),\vs^{-1})\right)\right|_{\vs=0},
\\
\label{C}
C_k
&=&
\frac{1}{(M-k)!}\left.\left(\frac{d^{M-k}}{d\ve^{M-k}}
g(H(\ve),\ve)\right)\right|_{\ve=0},
\ee
where $f(X,\lambda)$ and $g(X,\lambda)$ are properly defined functions of
operator $X$ and parameter $\lambda$.
The operator at the RHS of the first equation of the Lax pair now reads
\be\label{A}
\displaystyle A(\lambda)
&=&\sum_{k=0}^{L}\frac{\lambda^{L-k}}{k!}
\left.\left(\frac{d^k}{d\vs^k}f(\vs^NH(\vs^{-1}),\vs^{-
1})\right)\right|_{\vs=0}\nonumber\\
&\pp =&+
\sum_{k=0}^{M-1}\frac{\lambda^{k-
M}}{k!}\left.\left(\frac{d^k}{d\ve^k}g(H(\ve),\ve)
\right)\right|_{\ve=0}.
\ee
There exists a large class of functions $f(X,\lambda)$
and $g(X,\lambda)$ that results in operators $B_k$ and $C_k$ {\it
identically\/} satisfying conditions (\ref{setB}) and (\ref{setC}).
The class is defined by
\be
[f(X(\lambda),\lambda),X(\lambda)]=[g(X(\lambda),\lambda),X(\lambda)]=0.
\ee
To prove the covariance of Eqs.~(\ref{B}), (\ref{C}) under the
binary Darboux transformation we also assume that these functions
possess an additional property, namely they are covariant with
respect to the similarity
transformation:
\be\label{fg}
f(TXT^{-1},\lambda)=Tf(X,\lambda)T^{-1},\quad
g(TXT^{-1},\lambda)=Tg(X,\lambda)T^{-1},
\ee
where $T$ is a transformation.
The above conditions are satisfied, for example, by polynomials in $X$ and
sums of negative powers of polynomials in $X$.
If $X$ is selfadjoint, the same is valid for all $f(X)$ defined via the
spectral theorem.

In such a case Eqs.~(\ref{setB}) and (\ref{setC}) turn out to be identically
fulfilled as a consequence of the trivial identities
$$
\frac{d^n}{d\lambda^n}[f(H(\lambda),\lambda),H(\lambda)]\Big|_{\lambda=0}
\equiv 0,\quad
\frac{d^n}{d\lambda^n}
[g(H(\lambda),\lambda),H(\lambda)]\Big|_{\lambda=0}
\equiv 0
$$
and Eq.~(\ref{ode}) can be written in equivalent form
\be\label{eode}
i\dot H_m=\sum_{k=m+1}^{N}[H_k,B_{m-k}]+
\sum_{k=0}^m[H_k,C_{k-m}] \quad (0\le m\le N).
\ee
In the next section we will see that there are two representations of the
compatibility condition and they correspond to Eqs.~(\ref{Heis}) and
(\ref{vNeu}).

\section{Examples}

Below we present a few examples of integrable nonlinear von~Neumann-type
equations that correspond to different choices of positive integers $N$, $L$,
$M$ and functions $f(X,\lambda)$, $g(X,\lambda)$.
In what follows we will use the notation
$$
H_{1}=H,\quad H_{0}=\rho.
$$
If $N=1$ Eqs.~(\ref{ode}) imply
$$
\dot H=0.
$$

\subsection{}
\noindent
$N=1$, $L=1$, $f(X,\lambda)=X^n$ ($n\in {\bf N}$), $g(X,\lambda)=0$.
\medskip

The compatibility condition gives the equation
\be\label{X^n}
i\dot\rho=\left[\,\sum_{k=0}^{n-1}H^{n-k-1}\rho H^k,\rho\right].
\ee
The Darboux-covariant Lax pair for this equation was found in \cite{MKMCSL}.
For $n=2$ Eq.~(\ref{X^n}) reads
\be\label{X^2}
i\dot\rho=\left[\rho H+H\rho,\rho\right]=\left[H,\rho^2\right],
\ee
which is equivalent to Eq.~(\ref{EA}).
Mutual replacement of $H(\lambda)=\rho+\lambda H$ and $
A(\lambda)=H\rho+\rho H+\lambda H^2$ in the corresponding Lax pair results in
the compatibility condition
$$
i(H\dot\rho+\dot\rho H)=[H,\rho^2],
$$
which is essentially a form of Euler's top equations given in \cite{A,Mishch}.

\subsection{}
\noindent
$N=1$, $f(X,\lambda)=0$, $M=1$, $g(X,\lambda)=g(X)$.
\medskip

Here we have
$$
i\dot\rho=\left[g(\rho),H\right].
$$
The Lax-pair representation and Darboux covariance properties of this
equation have already been established in \cite{ULCK}.
It should be stressed that the function $g(X)$ is basically arbitrary.
The cases $g(\rho)=i\rho^3$ and $g(\rho)=i\rho^{-1}$ were considered in
\cite{MS}.

\subsection{}
\noindent
$N=1$, $L=3$,
$f(X,\lambda)=\lambda^{-2}(a_0X^2+(b_0+\lambda b_1)X^3+
(c_0+\lambda c_1+\lambda^2c_2)X^4)$, $g(X,\lambda)=0$.
\medskip

The compatibility condition becomes
\be
i\dot\rho &=&[h(\rho),\rho]=[H,F(\rho)],
\ee
where
\be
h(\rho)&=&
a_0(\rho H+H\rho)+b_0(\rho H^2+H\rho H+H^2\rho)+
b_1(\rho^2H+\rho H\rho+H\rho^2)
\nonumber\\
&\pp =&
+c_0(\rho H^3+H\rho H^2+H^2\rho H+H^3\rho)
\nonumber\\
&\pp =&
+
c_1(\rho^2H^2+\rho H\rho H+\rho H^2\rho+H\rho^2 H+H\rho H\rho+H^2\rho^2)
\nonumber\\
&\pp =&+
c_2(\rho^3 H+\rho^2H\rho+\rho H\rho^2+H\rho^3),
\nonumber\\
F(\rho)
&=&
a_0\rho^2+b_0(\rho^2H+\rho H\rho+H\rho^2)+b_1\rho^3
\nonumber\\
&\pp =&+
c_0(\rho^2 H^2+\rho H\rho H+\rho H^2\rho+
H\rho^2 H+H\rho H\rho+H^2\rho^2)
\nonumber\\
&\pp =&+
c_1(\rho^3 H+\rho^2 H\rho+\rho H\rho^2+H\rho^3)+c_2\rho^4.
\nonumber
\ee
Here $a_0$, $b_0$, $b_1$, $c_0$, $c_1$, $c_2$ are arbitrary complex
parameters independent of $\lambda$.
If the dot is a derivative with respect to a time variable $t$, they
can depend on $t$.
The same is also valid for the next example.
Let us note that the map $\rho\mapsto h(\rho)$ is not a function of $\rho$ in
the standard sense of the spectral theory \cite{Thirring}
(such as $g(\rho)$ of the previous subsection).
In particular, $[h(\rho),\rho]\neq 0$. We refer to such maps as
{\it non-Abelian functions\/}, or {\it non-Abelian nonlinearities\/}.
\subsection{}
\noindent
$N=1$, $f(X,\lambda)=0$, $M=2$,

\noindent
$g(X,\lambda)=(a_0+\lambda a_1)\left((b_0+\lambda b_1){\bf 1}+X\right)^{-
1}+(c_0+\lambda c_1)\big((d_0+\lambda d_1){\bf 1}+X\big)^{-1}
\big((e_0+\lambda e_1){\bf 1}+X\big)^{-1}$.
\medskip

In this case we obtain
\be
i\dot\rho &=&[H,F(\rho)],
\ee
where
\be
F(\rho)
&=&
a_0(b_0{\bf 1}+\rho)^{-1}(b_1{\bf 1}+H)
(b_0{\bf 1}+\rho)^{-1}-a_1(b_0{\bf 1}+\rho)^{-1}
\nonumber\\
&\pp =&
+c_0((d_0{\bf 1}+\rho)^{-1}(d_1{\bf 1}+H)(d_0{\bf 1}+\rho)^{-1}
(e_0{\bf 1}+\rho)^{-1}
\nonumber\\
&\pp =&
+(d_0{\bf 1}+\rho)^{-1}
(e_0{\bf 1}+\rho)^{-1}(e_1{\bf 1}+H)(e_0{\bf 1}+\rho)^{-1})
\nonumber\\
&\pp =&
-c_1(d_0{\bf 1}+\rho)^{-1}(e_0{\bf 1}+\rho)^{-1}.
\ee
As opposed to the previous examples $F(\rho)$ is a non-Abelian nonpolynomial
function.

For $N=1$ the nonlinear equations involve only two types of operators: $\rho$
and $H$.
Increasing $N$ we can introduce non-Abelian nonlinearities involving an
arbitrary number of different operators.

\subsection{}
\noindent
$N=2$, $L=2$, $f(X,\lambda)=X^2$, $g(X,\lambda)=0$.
\medskip

This is the simplest example of $N=2$ nonlinearity.
The compatibility conditions are
\be
i\dot\rho
&=&
[H^2,\rho]+[H_2,\rho^2]=[H^2+H_2\rho+\rho H_2,\rho],\nonumber\\
i\dot H
&=&[H_2,H\rho+\rho H],\nonumber\\
\dot H_2&=&0.
\ee
This system is equivalent to a nonlinear von Neumann equation with
two types of nonlinearity: One given in an implicit form and the other of
the Euler type.

\subsection{}
\noindent
$N=2$, $L=1$, $f(X,\lambda)=X$, $g(X,\lambda)=0$.

The Lax pair is
\be
\label{DCLP1}
z_\lambda\psi_\lambda
&=&
\psi_\lambda
\big(
H_0+\lambda H_1+\lambda^2
H_2\big),
\\
\label{DCLP2}
-i\dot\psi_\lambda
&=&
\psi_\lambda
\big(H_1+\lambda H_2\big)
\ee
with the compatibility conditions
\be
\dot H_2 &=& 0,\label{cc1}\\
i\dot H_1 &=& [H_2,H_0],\label{cc2}\\
i\dot H_0 &=& [H_1,H_0].\label{cc3}
\ee
Defining
\be
F_1&=&(H_0-H_2)/(2i),\\
F_2&=&(H_0+H_2)/2,\\
F_3&=&H_1/(2i)
\ee
and the connection $\nabla f=\dot f+[f,F_3]$ we can write
the compatibility conditions as
\be
\nabla F_1\label{F1}
&=&
i[F_2,F_3],
\\
\nabla F_2\label{F2}
&=&
i[F_3,F_1],\\
\nabla F_3 &=& i[F_1,F_2].\label{F3}
\ee
The connection can be trivialized if we find an invertible solution
$\xi$ of the linear problem
\be
\dot\xi &=& -\xi F_3.\label{xiF}
\ee
Then
\be
f_k=\xi F_k \xi^{-1}
\ee
satisfies the standard Nahm equations
\be
\dot f_1\label{N1}
&=&
i[f_2,f_3],
\\
\dot f_2
&=&
i[f_3,f_1],\label{N2}
\\
\dot f_3
&=&
i[f_1,f_2].\label{N3}
\ee

\section{Binary Darboux transformation}
The first step towards extending the  technique
of Darboux transformations to
integrable nonlinear von~Neumann-type equations on associative rings is to
establish the Darboux covariance of the Lax pair (\ref{Lax}) without
any additional constraints.
In this section we show that an appropriate formulation of the binary Darboux
transformation makes the Lax pair covariant.

Assume $\chi$ is a solution of the Lax pair with parameter $\nu$:
\be
\left\{
\begin{array}{rcl}
-i\dot\chi&=&\chi A(\nu)\\
z_{\nu}\chi&=&\chi H(\nu)
\end{array}
\right.
\ee
and $\varphi$ is a solution of the dual Lax pair with parameter $\mu$:
\be
\left\{
\begin{array}{rcl}
i\dot\varphi&=&A(\mu)\varphi\\
z_{\mu}\varphi&=&H(\mu)\varphi
\end{array}
\right. .
\ee
We further suppose that these systems can be related with an operator $P$
satisfying $P^2=P$ and
\be\label{P_t}
-i\dot P=PA(\nu)P_{\perp}-P_{\perp}A(\mu)P,
\ee
where $P_{\perp}={\bf 1}-P$.
The above assumptions are fulfilled, for example, if $\chi=\langle\chi|$ and
$\varphi=|\varphi\rangle$ are some ``bra" and ``ket" associated with
a Hilbert  space.
Then
$$
P=\frac{|\varphi\rangle\langle\chi|}{\langle\chi|\varphi\rangle}.
$$
Another example is provided by $m\times n$ matrix $\chi$ and
$n\times m$ matrix $\varphi$.
$P$ is then defined by
$$
P=\varphi(\chi\varphi)^{-1}\chi.
$$
Some realizations of Darboux transformations in infinite dimensional
cases were given in \cite{M-Salle}. Very recently a new construction of
the Darboux transformation in terms of Clifford numbers was described
in \cite{Clifford}.

Defining
\be\label{DT}
\psi[1]&=&\psi D_{\lambda},\\
D_{\lambda}&=&{\bf 1}+\frac{\nu-\mu}{\mu-\lambda}P\label{D}
\ee
we come to the following

\noindent
{\bf Theorem 1.}
{\em The Lax pair {\em (\ref{Lax})} with the coefficients defined by
Eqs.~{\em (\ref{ABC})}, {\em (\ref{H})} is covariant with respect to
the binary Darboux transformation
$\{\psi,A(\lambda),H(\lambda)\}\to\{\psi[1],A(\lambda)[1],H(\lambda)[1]\}$,
where
\be\label{TA}
A(\lambda)[1]&=&\sum_{k=0}^L\lambda^kB_k[1]+
\sum_{k=1}^M\frac{1}{\lambda^k}C_k[1],
\\
\label{TB}
B_k[1]&=&B_k+(\mu-\nu)\sum_{m=k+1}^L\left(\mu^{m-k-1}P_{\perp}B_mP-
\nu^{m-k-1}PB_mP_{\perp}\right)\\
\label{TC}
C_k[1]&=&C_k-(\mu-\nu)\sum_{m=k}^M\left(\mu^{k-m-1}P_{\perp}C_mP-
\nu^{k-m-1}PC_mP_{\perp}\right)
\ee
and}
\be\label{TH}
H(\lambda)[1]&=&\sum_{k=0}^N\lambda^kH_k[1],\\
\label{TH_k}
H_k[1]&=&H_k\nonumber\\
&\pp =&
+(\mu-\nu)\sum_{m=k+1}^N\left(\mu^{m-k-1}P_{\perp}H_mP-
\nu^{m-k-1}PH_mP_{\perp}\right)
\ee
{\bf Proof:}
The condition of covariance of the second equation of the Lax pair with
respect to the transformation yields
\be
H(\lambda)[1]&=&
D_{\lambda}^{-1}H(\lambda)D_{\lambda}\label{Chi}\\
&=&
\left({\bf 1}+\frac{\mu-\nu}{\nu-\lambda}P\right)
\sum_{k=0}^N\lambda^kH_k\left({\bf 1}
+\frac{\nu-\mu}{\mu-\lambda}P\right) \nonumber\\
&=&
\sum_{k=0}^N\lambda^kH_k+\frac{\mu-\nu}{\nu-\lambda}\sum_{k=0}^N\lambda^k
PH_kP_{\perp}+\frac{\nu-\mu}{\mu-\lambda}\sum_{k=0}^N\lambda^kP_{\perp}H_kP.
\nonumber
\ee
Taking into account
$$
PH(\nu)P_{\perp}=P_{\perp}H(\mu)P=0
$$
we are able to rewrite the previous expression in the following manner
\be
H(\lambda)[1]
&=&
\sum_{k=0}^N\lambda^kH_k+
\frac{\mu-\nu}{\nu-\lambda}\sum_{k=0}^N(\lambda^k-\nu^k)PH_kP_{\perp}
\nonumber\\
&\pp =&
+
\frac{\nu-\mu}{\mu-\lambda}\sum_{k=0}^N(\lambda^k-\mu^k)P_{\perp}H_kP
\nonumber\\
&=&
\sum_{k=0}^N\lambda^kH_k+
(\nu-\mu)\sum_{k=1}^N\sum_{j=0}^{k-1}\lambda^{k-j-1}\nu^jPH_kP_{\perp}
\nonumber\\
&\pp =&
+
(\mu-\nu)\sum_{k=1}^N\sum_{j=0}^{k-1}\lambda^{k-j-1}\mu^jP_{\perp}H_kP,
\ee
which is equivalent to Eq.~(\ref{TH}).

From the condition of the Darboux covariance of the first equation of the Lax
pair we have
$$
A(\lambda)[1]=D_{\lambda}^{-1}A(\lambda)D_{\lambda}-
iD_{\lambda}^{-1}\dot D_{\lambda}.
$$
Substitution of Eqs.~(\ref{P_t}), (\ref{D}) gives
\be
A(\lambda)[1]&=&\Big({\bf 1}+\frac{\mu-\nu}{\nu-\lambda}P\Big)
A(\lambda)\Big({\bf 1}+\frac{\nu-\mu}{\mu-\lambda}P\Big)
\nonumber\\
&\pp =&
+
\frac{\nu-\mu}{\mu-\lambda}\Big({\bf 1}+\frac{\mu-\nu}{\nu-\lambda}P\Big)
\big(PA(\nu)P_{\perp}-P_{\perp}A(\mu)P\big)
\nonumber\\
&=&
A(\lambda)+\frac{\mu-\nu}{\nu-\lambda}P(A(\lambda)-A(\nu))P_{\perp}
+\frac{\nu-\mu}{\mu-\lambda}P_{\perp}(A(\lambda)-A(\mu))P\nonumber\\
&=&
\sum_{k=0}^L\lambda^kB_k+\sum_{k=1}^M\frac{1}{\lambda^k}C_k
\nonumber\\
&\pp =&
+
(\mu-\nu)P_{\perp}\left(\sum_{k=1}^L\sum_{j=0}^{k-1}\lambda^{k-j-1}\mu^jB_k
-\sum_{k=1}^M\sum_{j=0}^{k-1}\lambda^{-j-1}\mu^{j-k}C_k\right)P
\nonumber\\
&\pp =&
+(\nu-\mu)P\left(\sum_{k=1}^L\sum_{j=0}^{k-1}\lambda^{k-j-1}\nu^jB_k
-\sum_{k=1}^M\sum_{j=0}^{k-1}\lambda^{-j-1}\nu^{j-k}C_k\right)P_{\perp}.
\nonumber
\ee
The final expression is (\ref{TA}).%
\rule{5pt}{5pt}

\section{The main theorem}

Theorem~1 establishes Darboux covariance of Lax pairs
involving operators with positive or negative powers of spectral parameters.
The compatibility condition is also covariant: Transformed operators
$B_k[1]$,
$C_k[1]$ and $H_k[1]$ solve Eqs.~(\ref{setB}), (\ref{setC}), and (\ref{ode}).
However, if there are additional relations between the operators, they do not
have to be Darboux covariant.
\medskip

\noindent
{\bf Theorem 2.} {\em The relations {\em (\ref{B})} and {\em (\ref{C})} are
Darboux covariant if Eqs.~{\em (\ref{fg})} are fulfilled.\/}

\noindent
{\bf Proof:}
Let us check the Darboux covariance of Eq.~(\ref{B}), i.e.
$$
B_k[1]=\frac{1}{(L-k)!}\left.\left(\frac{d^{L-k}}{d\vs^{L-k}}
f(\vs^NH(\vs^{-1})[1],\vs^{-1})\right)\right|_{\vs=0}.
$$
It follows immediately that
$$
B_L[1]=B_L
$$
and
\be
\frac{d^k}{d\vs^k}D_{1/\vs}\Big|_{\vs=0}
&=&
k!(\mu-\nu)\mu^{k-1}P,\nonumber\\
\frac{d^k}{d\vs^k}D_{1/\vs}^{-1}\Big|_{\vs=0}
&=&
k!(\nu-\mu)\nu^{k-1}P.\nonumber
\ee
For $k\ne L$ we have, using Eq.~(\ref{Chi}),
\be
{}&{}&
\frac{1}{(L-k)!}\left.\left(\frac{d^{L-k}}{d\vs^{L-k}}
f(\vs^NH(\vs^{-1})[1],\vs^{-1})\right)\right|_{\vs=0}
\nonumber\\
&{}&
=
\frac{1}{(L-k)!}\left.\left(\frac{d^{L-k}}{d\vs^{L-k}}f(\vs^ND_{1/\vs}^{-1}
H(\vs^{-1})D_{1/\vs},\vs^{-1})\right)\right|_{\vs=0}\nonumber\\
&{}&
=\frac{1}{(L-k)!}
\left.\left(\frac{d^{L-k}}{d\vs^{L-k}}\left(D_{1/\vs}^{-1}
f(\vs^NH(\vs^{-1}),\vs^{-1})D_{1/\vs}\right)\right)\right|_{\vs=0}
\nonumber\\
&{}&
=\frac{1}{(L-k)!}
\sum_{a=0}^{L-k}\sum_{b=0}^{a}\frac{(L-k)!}{(L-k-a)!a!}\frac{a!}{(a-b)!b!}
\nonumber\\
&{}&
\times
\left.\left(\frac{d^{L-k-a}}{d\vs^{L-k-a}}D_{1/\vs}^{-1}
\right)\right|_{\vs=0}
\left.\left(\frac{d^{a-b}}{d\vs^{a-b}}f(\vs^NH(\vs^{-1}),\vs^{-1})\right)
\right|_{\vs=0}\left.\left(\frac{d^b}{d\vs^b}D_{1/\vs}\right)
\right|_{\vs=0}\nonumber\\
&{}&
=\sum_{a=0}^{L-k}\sum_{b=0}^{a}\frac{1}{(L-k-a)!b!}
\left.\left(\frac{d^{L-k-a}}{d\vs^{L-k-a}}D_{1/\vs}^{-1}\right)
\right|_{\vs=0}B_{L+b-a}\left.\left(\frac{d^b}{d\vs^b}D_{1/\vs}
\right)\right|_{\vs=0}\nonumber\\
&{}&
=\sum_{a=1}^{L-k}\sum_{b=0}^{a}\frac{1}{(L-k-a)!b!}
\left.\left(\frac{d^{L-k-a}}{d\vs^{L-k-a}}D_{1{/}\vs}^{-1}\right)
\right|_{\vs=0}B_{L+b-a}\left.\left(\frac{d^b}{d\vs^b}D_{1{/}\vs}
\right)\right|_{\vs=0}
\nonumber\\
&{}&
\pp {==}
+(\nu-\mu)\nu^{L-k-1}PB_L
\nonumber\\
&{}&
=
\sum_{a=1}^{L-k}\sum_{b=1}^{a}\frac{(\mu-\nu)\mu^{b-1}}{(L-k-a)!}
\left.\left(\frac{d^{L-k-a}}{d\vs^{L-k-a}}D_{1{/}\vs}^{-1}\right)
\right|_{\vs=0}B_{L+b-a}P\nonumber\\
&{}&\pp {==}
+\sum_{a=1}^{L-k}\frac{1}{(L-k-a)!}\left.\left(\frac{d^{L-k-a}}{d\vs^{L-k-a}}
D_{1{/}\vs}^{-1}\right)\right|_{\vs=0}B_{L-a}+(\nu-\mu)\nu^{L-k-1}PB_L
\nonumber\\
&{}&
=\sum_{a=1}^{L-k-1}\!\!\sum_{b=1}^{a}\frac{(\mu-\nu)\mu^{b-1}}{(L-k-a)!}
\left.\left(\frac{d^{L-k-a}}{d\vs^{L-k-a}}D_{1{/}\vs}^{-1}\right)
\right|_{\vs=0}B_{L+b-a}P
\nonumber\\
&{}&\pp {==}
+\sum_{b=1}^{L-k}(\mu-\nu)\mu^{b-1}B_{k+b}P
\nonumber\\
&{}&\pp {==}
+\sum_{a=1}^{L-k-1}\frac{1}{(L-k-a)!}\left.\left(\frac{d^{L-k-a}}
{d\vs^{L-k-a}}
D_{1{/}\vs}^{-1}\right)\right|_{\vs=0}B_{L-a}
\nonumber\\
&{}&\pp {==}
+B_k+(\nu-\mu)\nu^{L-k-1}PB_L
\nonumber\\
&{}&
=\sum_{a=1}^{L-k-1}\sum_{b=1}^{a}(\mu-\nu)\mu^{b-1}(\nu-\mu)\nu^{L-k-a-1}
PB_{L+b-a}P
\nonumber\\
&{}&\pp {==}
+\sum_{b=1}^{L-k}(\mu-\nu)\mu^{b-1}B_{k+b}P
\nonumber\\
&{}&\pp {==}
+\sum_{a=1}^{L-k-1}(\nu-\mu)\nu^{L-k-a-1}PB_{L-a}+B_k
+(\nu-\mu)\nu^{L-k-1}PB_L
\nonumber\\
&{}&
=B_k+(\mu-\nu)\left(\sum_{b=1}^{L-k}\mu^{b-1}B_{k+b}P-
\sum_{a=0}^{L-k-1}\nu^{L-k-a-1}PB_{L-a}\right.
\nonumber\\
&{}&
\pp{=B_k+(\mu-\nu)\sum_{b=1}^{L-k}}
+
\left.(\nu-\mu)\sum_{a=1}^{L-k-1}\sum_{b=1}^{a}\mu^{b-1}\nu^{L-k-a-1}PB_{L+b-
a}P\right)
\nonumber\\
&{}&
=B_k+(\mu-\nu)\left(\sum_{b=1}^{L-k}\mu^{b-1}P_{\perp}B_{k+b}P-
\sum_{a=0}^{L-k-1}\nu^{L-k-a-1}PB_{L-a}P_{\perp}+\delta_k\right),
\nonumber
\ee
where
\be
\delta_k
&=&
\sum_{b=1}^{L-k}\mu^{b-1}PB_{k+b}P-
\sum_{a=0}^{L-k-1}\nu^{L-k-a-1}PB_{L-a}P
\nonumber\\
&\pp =&
+
(\nu-\mu)\sum_{a=1}^{L-k-1}\sum_{b=1}^{a}\mu^{b-1}\nu^{L-k-a-1}PB_{L+b-a}P
\nonumber\\
&=&
\sum_{b=1}^{L-k}\mu^{b-1}PB_{k+b}P-
\sum_{a=0}^{L-k-1}\nu^{L-k-a-1}PB_{L-a}P
\nonumber\\
&\pp =&
+\sum_{a=1}^{L-k-1}\sum_{b=1}^{a}\mu^{b-1}\nu^{L-k-a}PB_{L+b-a}P
\nonumber\\
&&
-
\sum_{a=1}^{L-k-1}\sum_{b=1}^{a}\mu^b\nu^{L-k-a-1}PB_{L+b-a}P.
\nonumber
\ee
Combining, respectively, the first and the third, the second and the
fourth terms gives
$$
\delta_k=\sum_{a=1}^{L-k}\sum_{b=1}^{a}\mu^{b-1}\nu^{L-k-a}PB_{L+b-a}P-\!\!
\sum_{a=0}^{L-k-1}\!\sum_{b=0}^{a}\mu^b\nu^{L-k-a-1}PB_{L+b-a}P\equiv0.
$$
Finally, we obtain
\be
{}&{}&
\!\!\!\!\!\!\!\!\!\frac{1}{(L-k)!}\left.\left(\frac{d^{L-k}}{d\vs^{L-k}}
f(\vs^NH(\vs^{-1})[1],\vs^{-1})\right)\right|_{\vs=0}
\nonumber\\
&{}&
\!\!\!\!\!\!=
B_k+(\mu-\nu)\left(\sum_{m=1}^{L-k}\mu^{m-1}P_{\perp}B_{k+m}P-\!\!
\sum_{m=0}^{L-k-1}\!\nu^{L-k-m-1}PB_{L-m}P_{\perp}\right).
\ee
The last expression coincides with Eq.~(\ref{TB}).

Let us prove the Darboux covariance of Eq.~(\ref{C}):
$$
C_k[1]=\frac{1}{(M-k)!}\left.\left(\frac{d^{M-k}}{d\ve^{M-k}}
g(H(\ve)[1],\ve)\right)\right|_{\ve=0}.
$$
One can show that
\be
\frac{d^k}{d\ve^k}D_{\ve}\Big|_{\ve=0}
&=&
\delta_{k0}{\bf 1}+k!(\nu-\mu)\mu^{-k-1}P,\nonumber\\
\frac{d^k}{d\ve^k}D_{\ve}^{-1}\Big|_{\ve=0}
&=&
\delta_{k0}{\bf 1}+k!(\mu-\nu)\nu^{-k-1}P.
\ee
Then
\be
{}&{}&
\frac{1}{(M-k)!}\left.\left(\frac{d^{M-k}}{d\ve^{M-k}}
g(H(\ve)[1],\ve)\right)\right|_{\ve=0}
\nonumber\\
&{}&
=
\frac{1}{(M-k)!}
\left.\left(\frac{d^{M-k}}{d\ve^{M-k}}g(D_{\ve}^{-1}H(\ve)D_{\ve},\ve)
\right)\right|_{\ve=0}
\nonumber\\
&{}&
=\frac{1}{(M-k)!}
\left.\left(\frac{d^{M-k}}{d\ve^{M-k}}\left(D_{\ve}^{-
1}g(H(\ve),\ve)D_{\ve}\right)
\right)\right|_{\ve=0}
\nonumber\\
&{}&
=\frac{1}{(M-k)!}
\sum_{a=0}^{M-k}\sum_{b=0}^{a}\frac{(M-k)!}{(M-k-a)!a!}\frac{a!}{(a-b)!b!}
\nonumber\\
&{}&
\times
\left.\left(\frac{d^{M-k-a}}{d\ve^{M-k-a}}D_{\ve}^{-1}\right)\right|_{\ve=0}
\left.\left(\frac{d^{a-b}}{d\ve^{a-b}}g(H(\ve),\ve)\right)\right|_{\ve=0}
\left.\left(\frac{d^b}{d\ve^b}D_{\ve}\right)\right|_{\ve=0}
\nonumber\\
&{}&
=\sum_{a=0}^{M-k}\sum_{b=0}^{a}\frac{1}{(M-k-a)!b!}
\nonumber\\
&{}&\pp{==}
\times
\Big(\delta_{(M-k-a)0}{\bf 1}
+(M-k-a)!(\mu-\nu)\nu^{k+a-M-1a}P\Big)
\nonumber\\
&{}&\pp{==}
\times
C_{M+b-a}\Big(\delta_{b0}{\bf 1}+b!(\nu-\mu)\mu^{-b-1}P\Big)
\nonumber\\
&{}&
=\left({\bf 1}+\frac{\mu-\nu}{\nu}P\right)C_k
\left({\bf 1}
+\frac{\nu-\mu}{\mu}P\right)\nonumber\\
&{}&\pp {==}
+
(\nu-\mu)\sum_{b=1}^{M-k}\mu^{-b-1}
\left({\bf 1}+\frac{\mu-\nu}{\nu}P\right)C_{b+k}P\nonumber\\
&{}&\pp {==}
+(\mu-\nu)\sum_{a=0}^{M-k-1}
\nu^{k+a-M-1}PC_{M-a}\left({\bf 1}+\frac{\nu-\mu}{\mu}P\right)
\nonumber\\
&{}&\pp {==}
-
(\mu-\nu)^2\sum_{a=1}^{M-k-1}\sum_{b=1}^{a}\nu^{k+a-M-1}\mu^{-b-1}
PC_{M+b-a}P
\nonumber\\
&{}&
=\left({\bf 1}+\frac{\mu-\nu}{\nu}P\right)C_k
\left({\bf 1}+\frac{\nu-\mu}{\mu}P\right)
\nonumber\\
&{}&
+
(\mu-\nu)\left(\sum_{a=0}^{M-k-1}\!\nu^{k+a-M-1}PC_{M-a}P_{\perp}-
\sum_{b=1}^{M-k}\mu^{-b-1}P_{\perp}C_{b+k}P+\Delta_k\right),\nonumber
\ee
where
\be
\Delta_k
&=&
\sum_{a=0}^{M-k-1}\nu^{k+a-M}\mu^{-1}PC_{M-a}P-
\sum_{b=1}^{M-k}\mu^{-b}\nu^{-1}PC_{b+k}P
\nonumber\\
&\pp =&\pp =
+
(\nu-\mu)\sum_{a=1}^{M-k-1}\sum_{b=1}^{a}\mu^{-b-1}\nu^{k+a-M-1}PC_{M+b-a}P
\nonumber\\
&=&\sum_{a=0}^{M-k-1}\nu^{k+a-M}\mu^{-1}PC_{M-a}P-
\sum_{b=1}^{M-k}\mu^{-b}\nu^{-1}PC_{b+k}P
\nonumber\\
&{}&\pp{==}
+\sum_{a=1}^{M-k-1}\sum_{b=1}^{a}\mu^{-b-1}\nu^{k+a-M}PC_{M+b-a}P
\nonumber\\
&{}&\pp{==}
-
\sum_{a=1}^{M-k-1}\sum_{b=1}^{a}\mu^{-b}\nu^{k+a-M-1}PC_{M+b-a}P.
\nonumber
\ee
Combining, respectively, the first and the third, the second and the
fourth terms we obtain
\be
\Delta_k&=&\sum_{a=0}^{M-k-1}\sum_{b=0}^{a}\mu^{-b-1}\nu^{k+a-M}
PC_{M+b-a}P
\nonumber\\
&\pp =&
-\sum_{a=1}^{M-k}\sum_{b=1}^{a}\mu^{-b}\nu^{k+a-M-1}PC_{M+b-a}P
\equiv 0.\nonumber
\ee
Finally, we have
\be
{}&{}&
\frac{1}{(M-k)!}\left.\left(\frac{d^{M-k}}{d\ve^{M-k}}
g(H(\ve)[1],\ve)\right)\right|_{\ve=0}
\nonumber\\
&{}&\pp =
=
\left({\bf 1}+\frac{\mu-\nu}{\nu}P\right)C_k
\left({\bf 1}+\frac{\nu-\mu}{\mu}P\right)
\nonumber\\
&{}&\pp{==}
+(\mu-\nu)\left(\sum_{a=0}^{M-k-1}\nu^{k+a-M-1}PC_{M-a}P_{\perp}-
\sum_{b=1}^{M-k}\mu^{-b-1}P_{\perp}C_{b+k}P\right)
\nonumber\\
&{}&\pp =
=
C_k+\frac{\mu-\nu}{\nu}PC_kP_{\perp}+\frac{\nu-\mu}{\mu}P_{\perp}C_kP
\nonumber\\
&{}&\pp{==}
+(\mu-\nu)\left(\sum_{a=0}^{M-k-1}\nu^{k+a-M-1}PC_{M-a}P_{\perp}-
\sum_{b=1}^{M-k}\mu^{-b-1}P_{\perp}C_{b+k}P\right)
\nonumber\\
&{}&\pp =
=C_k+(\mu-\nu)\left(\sum_{a=0}^{M-k}\nu^{k+a-M-1}PC_{M-a}P_{\perp}
-\sum_{b=0}^{M-k}\mu^{-b-1}P_{\perp}C_{b+k}P\right).
\nonumber
\ee
The last expression coincides with Eq.~(\ref{TC}).%
\rule{5pt}{5pt}

\section{Conclusions}

We have established the Darboux-covariance of a large class of
nonlinear von Neumann-type equations. The next step is to employ this
fact in construction of explicit solutions of such equations.
Some classes of solutions have already been found in
\cite{SLMC,ULCK,CKLN} for nonlinearities
\be
i\dot \rho=[H,f(\rho)]
\ee
with $f(\rho)=\rho^2$ and $f(\rho)=\rho^q-2\rho^{q-1}$ ($q$ is an
arbitrary real number).
Both finite- and infinite-dimensional cases were treated by this technique
in \cite{ULCK}.

In a forthcoming paper we will describe other classes of solutions of
the integrable equations we have introduced.
It also seems that Lax pairs that
allow us to reduce the compatibility conditions to nonlinear
equations in a closed form can be still generalized.
We hope in the future work to develop a description of integrable
equations of the von~Neumann type by taking into consideration the
Mikhailov method of automorphisms \cite{Mikh}. This type of
generalization is particularly important if reductions characteristic
of Nahm-type equations are involved.

\section*{Acknowledgments}
M.C. is indebted to Jan L. Cie\'sli\'nski for his comments and, in
particular, for the suggestion of using the similarity-transformation
form of the Darboux transformation. The work of M.C. was supported by the
Alexander von Humboldt Foundation and the KBN Grant 5 P03B 040 20.
The work of N.V.U. was supported by Nokia-Poland.

\end{document}